\documentclass{article}
\usepackage{amssymb}
\usepackage{amsfonts}
\usepackage{amsmath}

\input{tcilatex}

\begin{document}

\begin{center}
\bigskip

{\large A RENORMALIZABLE QUANTUM\ FIELD\ THEORETIC\ MODEL WITH\ GRAVITY}

{\large \ }\bigskip

C. N. Ragiadakos

IEP, Tsoha 36, Athens 11521, Greece

email: crag@iep.edu.gr

\bigskip

Essay written for the Gravity Research Foundation 2012 Awards for Essays on
Gravitation

\bigskip

\textbf{ABSTRACT}
\end{center}

\begin{quote}
A four dimensional generally covariant modified Yang-Mills action, which
depends on the lorentzian complex structure of spacetime and not its metric,
is presented. The extended Weyl symmetry, implied by the effective metric
independence, makes the lagrangian model renormalizable. The modified
Yang-Mills action generates a linear potential, instead of the Coulomb-like $%
\frac{1}{r}$ potential of the ordinary action. Therefore the Yang-Mills
excitations must be perturbatively confined. The metric, which admits an
integrable lorentzian complex structure, can be extended to a Kaehler metric
and the spacetime is a totally real CR manifold in $%
%TCIMACRO{\U{2102} }%
%BeginExpansion
\mathbb{C}
%EndExpansion
^{4}$. These surfaces are generally inside the SU(2,2) homogeneous domain. A
non-real-analytic point, transferred to the U(2) characteristic boundary of
the classical domain, spontaneously breaks the SU(2,2) symmetry down to its
Poincar\'{e} subgroup. Hence the pure geometric modes and solitons of the
model must belong to representations of the Poincar\'{e} group.

\newpage 
\end{quote}

\setcounter{equation}{0}

The linearized string action describes the dynamics of 2-dimensional
surfaces in a multidimensional space. Its form%
\begin{equation}
I_{S}=\frac{1}{2}\int d^{2}\!\xi \ \sqrt{-\gamma }\ \gamma ^{\alpha \beta }\
\partial _{\alpha }X^{\mu }\partial _{\beta }X^{\nu }\eta _{\mu \nu }
\label{e1}
\end{equation}%
does not essentially depend on the metric $\gamma ^{\alpha \beta }$ of the
2-dimensional surface but on its complex structure. It depends on its
structure coordinates $(z^{0},\ z^{\widetilde{0}})$, because in these
coordinates it takes the metric independent form%
\begin{equation}
I_{S}=\int d^{2}\!z\ \partial _{0}X^{\mu }\partial _{\widetilde{0}}X^{\nu
}\eta _{\mu \nu }  \label{e2}
\end{equation}%
All the wonderful properties of the string model are essentially based on
this characteristic feature of the string action.

The plausible question\cite{RAG1988} and exercise is \textquotedblleft what
4-dimensional action with first order derivatives depends on the complex
structure but it does not depend on the metric of the
spacetime?\textquotedblright . The additional expectation is that such an
action may be formally renormalizable, because the regularization procedure
will not generate geometric counterterms. The term \textquotedblleft
formally\textquotedblright\ is used, because the 4-dimensional action may
have anomalies, which could destroy renormalizability, as it happens in the
string action.

The lorentzian signature of spacetime is not compatible\cite{FLAHE1974} with
a \underline{real} tensor (complex structure) $J_{\mu }^{\;\nu }$. Therefore
Flaherty introduced a complex tensor to define the Lorentzian complex
structure, which he extensively studied\cite{FLAHE1974}. It can be shown
that there is always a null tetrad $(\ell _{\mu },\,n_{\mu },\,m_{\mu },\,%
\overline{m}_{\mu })$ such that the metric tensor and the complex structure
tensor take the form

\begin{equation}
\begin{array}{l}
g_{\mu \nu }=\ell _{\mu }n_{\nu }+n_{\mu }\ell _{\nu }-m_{{}\mu }\overline{m}%
_{\nu }-\overline{m}_{\mu }m_{\nu } \\ 
\\ 
J_{\mu }^{\;\nu }=i(\ell _{\mu }n^{\nu }-n_{\mu }\ell ^{\nu }-m_{\mu }%
\overline{m}^{\nu }+\overline{m}_{\mu }m^{\nu })%
\end{array}
\label{e3}
\end{equation}%
The integrability condition of this complex structure implies the Frobenius
integrability conditions of the pairs $(\ell _{\mu },\,\,m_{\mu })$ and $%
(n_{\mu },\,\overline{m}_{\mu })$

\begin{equation}
\begin{array}{l}
(\ell ^{\mu }m^{\nu }-\ell ^{\nu }m^{\mu })(\partial _{\mu }\ell _{\nu
})=0\;\;\;\;,\;\;\;\;(\ell ^{\mu }m^{\nu }-\ell ^{\nu }m^{\mu })(\partial
_{\mu }m_{\nu })=0 \\ 
\\ 
(n^{\mu }m^{\nu }-n^{\nu }m^{\mu })(\partial _{\mu }n_{\nu
})=0\;\;\;\;,\;\;\;\;(n^{\mu }m^{\nu }-n^{\nu }m^{\mu })(\partial _{\mu
}m_{\nu })=0%
\end{array}
\label{e4}
\end{equation}%
That is, only metrics with two geodetic and shear free congruences ($\kappa
=\sigma =\lambda =\nu =0$)\cite{P-R1984} admit an integrable complex
structure.

Frobenius theorem states that there are four complex functions $%
z^{b}=(z^{\alpha },\;z^{\widetilde{\alpha }})$,\ $\alpha =0,\ 1$ , such that

\begin{equation}
dz^{\alpha }=f_{\alpha }\ \ell _{\mu }dx^{\mu }+h_{\alpha }\ m_{\mu }dx^{\mu
}\;\;\;\;,\;\;\;dz^{\widetilde{\alpha }}=f_{\widetilde{\alpha }}\ n_{\mu
}dx^{\mu }+h_{\widetilde{\alpha }}\ \overline{m}_{\mu }dx^{\mu }\;
\label{e5}
\end{equation}%
These four functions are the structure coordinates of the (integrable)
complex structure. Notice that in the present case of lorentzian spacetimes
the coordinates $z^{\widetilde{\alpha }}$ are not complex conjugate of $%
z^{\alpha }$, because $J_{\mu }^{\;\nu }$ is no longer a real tensor.

Using these structure coordinates, a metric independent action takes the
simple form 
\begin{equation}
\begin{array}{l}
I_{G}=\frac{1}{2}\int d^{4}\!z\det (g_{\alpha \widetilde{\alpha }})\
g^{\alpha \widetilde{\beta }}g^{\gamma \widetilde{\delta }}F_{\!j\alpha
\gamma }F_{\!j\widetilde{\beta }\widetilde{\delta }}+comp.\ conj.=\int
d^{4}\!z\ F_{\!j01}F_{\!j\widetilde{0}\widetilde{1}}+c.\ c. \\ 
\\ 
F_{j_{ab}}=\partial _{a}A_{jb}-\partial _{a}A_{jb}-\gamma
\,f_{jik}A_{ia}A_{kb}%
\end{array}
\label{e6}
\end{equation}%
This transcription is possible because the metric takes the simple form $%
ds^{2}=2g_{\alpha \widetilde{\beta }}dz^{\alpha }dz^{\widetilde{\beta }}$ in
the structure coordinates system.

The covariant null tetrad form of this action\cite{RAG1990} is 
\begin{equation}
\begin{array}{l}
I_{G}=\int d^{4}\!x\ \sqrt{-g}\ \left\{ \left( \ell ^{\mu }m^{\rho
}F_{\!j\mu \rho }\right) \left( n^{\nu }\overline{m}^{\sigma }F_{\!j\nu
\sigma }\right) +\left( \ell ^{\mu }\overline{m}^{\rho }F_{\!j\mu \rho
}\right) \left( n^{\nu }m^{\sigma }F_{\!j\nu \sigma }\right) \right\}  \\ 
\\ 
F_{j\mu \nu }=\partial _{\mu }A_{j\nu }-\partial _{\nu }A_{j\mu }-\gamma
\,f_{jik}A_{i\mu }A_{k\nu }%
\end{array}
\label{e7}
\end{equation}%
where $A_{j\mu }$ is an $SU(N)$ gauge field and $(\ell _{\mu },\,n_{\mu
},\,m_{\mu },\,\overline{m}_{\mu })$ is the special integrable null tetrad (%
\ref{e3}). The difference between the present action and the ordinary
Yang-Mills action becomes more clear in the following form of the action 
\begin{equation}
I_{G}=-\frac{1}{8}\int d^{4}\!x\ \sqrt{-g}\ \left( 2g^{\mu \nu }\ g^{\rho
\sigma }-J^{\mu \nu }\ J^{\rho \sigma }-\overline{J^{\mu \nu }}\ \overline{%
J^{\rho \sigma }}\right) F_{\!j\mu \rho }F_{\!j\nu \sigma }  \label{e8}
\end{equation}%
where $g_{\mu \nu }$ is a metric derived from the null tetrad (\ref{e3}) and 
$J_{\mu }^{\;\nu }$ is the corresponding tensor of the integrable complex
structure.

In the case of the string action (\ref{e1}) we do not need additional
conditions, because any orientable 2-dimensional surface admits a complex
structure. But in the case of 4-dimensional surfaces, the integrability of
the complex structure has to be imposed through precise conditions. These
integrability conditions (\ref{e4}) may be imposed using the ordinary
procedure of Lagrange multipliers 
\begin{equation}
\begin{array}{l}
I_{C}=\int d^{4}\!x\ \{\phi _{0}(\ell ^{\mu }m^{\nu }-\ell ^{\nu }m^{\mu
})(\partial _{\mu }\ell _{\nu })+ \\ 
\\ 
\qquad +\phi _{1}(\ell ^{\mu }m^{\nu }-\ell ^{\nu }m^{\mu })(\partial _{\mu
}m_{\nu })+\phi _{\widetilde{0}}(n^{\mu }\overline{m}^{\nu }-n^{\nu }%
\overline{m}^{\mu })(\partial _{\mu }n_{\nu })+ \\ 
\\ 
\qquad +\phi _{\widetilde{1}}(n^{\mu }\overline{m}^{\nu }-n^{\nu }\overline{m%
}^{\mu })(\partial _{\mu }\overline{m}_{\nu })+c.conj.\}%
\end{array}
\label{e9}
\end{equation}%
This technique makes the complete action $I=I_{G}+I_{C}$ self-consistent and
the usual quantization techniques may be applied\cite{RAG1992}.

The local symmetries of the action are a) the well known local gauge
transformations, b) the reparametrization symmetry as it is the case in any
generally covariant action and c) the following extended Weyl transformation
of the tetrad%
\begin{equation}
\begin{tabular}{l}
$\ell _{\mu }^{\prime }=\Lambda \ell _{\mu }\quad ,\quad \ell ^{\prime \mu }=%
\frac{1}{N}\ell ^{\mu }$ \\ 
\\ 
$n_{\mu }^{\prime }=Nn_{\mu }\quad ,\quad n^{\prime \mu }=\frac{1}{\Lambda }%
n^{\mu }$ \\ 
\\ 
$m_{\mu }^{\prime }=Mm_{\mu }\quad ,\quad m^{\prime \mu }=\frac{1}{\overline{%
M}}m^{\mu }$ \\ 
\end{tabular}
\label{e10}
\end{equation}%
which is larger than the ordinary Weyl (conformal) transformation.

A simple way\cite{RAG1991},\cite{RAG1999} to find a curved space complex
structure is the Kerr-Schild ansatz

\begin{equation}
\ell _{\mu }=L_{\mu }\quad ,\quad m_{\mu }=M_{\mu }\quad ,\quad n_{\mu
}=N_{\mu }+f(x)\ L_{\mu }  \label{e11}
\end{equation}%
where the null tetrad%
\begin{equation}
\begin{array}{l}
L_{\mu }dx^{\mu }=dt-dr-a\sin ^{2}\theta \ d\varphi \\ 
\\ 
N_{\mu }dx^{\mu }=\frac{r^{2}+a^{2}}{2(r^{2}+a^{2}\cos ^{2}\theta )}[dt+%
\frac{r^{2}+2a^{2}\cos ^{2}\theta -a^{2}}{r^{2}+a^{2}}dr-a\sin ^{2}\theta \
d\varphi ] \\ 
\\ 
M_{\mu }dx^{\mu }=\frac{-1}{\sqrt{2}(r+ia\cos \theta )}[-ia\sin \theta \
(dt-dr)+(r^{2}+a^{2}\cos ^{2}\theta )d\theta + \\ 
\qquad \qquad +i\sin \theta (r^{2}+a^{2})d\varphi ]%
\end{array}
\label{e12}
\end{equation}%
determines an integrable flat complex structure. In this case, $(\ell _{\mu
},\ n_{\mu },\ m_{\mu },\ \overline{m}_{\mu })$ is integrable for

\begin{equation}
f=\frac{h(r)}{2(r^{2}+a^{2}\cos ^{2}\theta )}  \label{e13}
\end{equation}%
where $h(r)$ is an arbitrary function. Notice that for $h(r)=-2mr+e^{2}$ the
Kerr-Newman space-time is found. A set of structure coordinates of this
curved complex structure is

\begin{equation}
\begin{array}{l}
z^{0}=t-r+ia\cos \theta -ia\quad ,\quad z^{1}=e^{i\varphi }\tan \frac{\theta 
}{2} \\ 
\\ 
z^{\widetilde{0}}=t+r-ia\cos \theta +ia-2f_{1}\quad ,\quad z^{\widetilde{1}%
}=-\frac{r+ia}{r-ia}\ e^{2iaf_{2}}\ e^{-i\varphi }\tan \frac{\theta }{2}%
\end{array}
\label{e14}
\end{equation}%
where the two functions are

\begin{equation}
f_{1}(r)=\int \frac{h}{r^{2}+a^{2}+h}\ dr\quad ,\quad f_{2}(r)=\int \frac{h}{%
(r^{2}+a^{2}+h)(r^{2}+a^{2})}\ dr  \label{e15}
\end{equation}%
Notice that spherically symmetric spacetimes ($a=0$) are compatible with the
trivial complex structure, which is compatible with the Minkowski metric too.

After the recent failure of ATLAS and CMS experiments to find natural
supersymmetry effects and (large) higher spacetime dimensions, doubts on the
physical relevance of the superstring model start to appear. Other road maps
for a unification of Quantum Theory and General Relativity have to be
investigated. Therefore I will actually outline the characteristic
properties of the present generally covariant lagrangian model which
indicate its possible physical relevance.

1)\ Renormalizability: The ordinary perturbative expansion of the lagrangian
around the trivial null tetrad\cite{RAG2008a} is dimensionless. Therefore
the number of counterterms is restricted. The extended Weyl symmetry (\ref%
{e10}) and the subsequent effective metric independence of the action
excludes any geometric counterterm, which makes ordinary Einstein action
non-renormalizable. Besides, my calculations\cite{RAG2008a} of the first
order 1-loop diagrams turn out to be finite! This is the cornerstone of the
present model. If a certain anomaly is found to destroy its formal
renormalizability, we have to abandon the model. The compatibility with
Quantum Mechanics is the crucial property that forces many particle
physicists to continue to work in the superstring model despite its negative
experimental results. They think that the string model is the unique known
self-consistent one with gravity, which is not true! The present model is
also self-consistent, while I do not actually know a third one.

2) Possible Existence of\ "Leptonic"\ and\ "Hadronic"\ Sectors: The
integrability conditions (\ref{e4}) decouple from the gauge field equations%
\cite{RAG2010}. Therefore the spectrum of the model with $A_{j\mu }=0$ will
not have any gauge field potential (charge). This spectrum consists of the
modes of the lorentzian complex structure and the possible geometric solitons%
\cite{RAG1999}. The solution of the linear part of the static gauge field
equations with an external source in the simple static complex structure (%
\ref{e12}) with $a=0$ is found to be linear, instead of the Coulomb-like $%
\frac{1}{r}$ potential of the ordinary Yang-Mills action. That is the $N$
gauge field excitations, which we will call "quarks", must be perturbatively
confined\cite{RAG2010}. Therefore the model may have two different sectors.
The "leptonic" sector with the pure geometric solitons (with $A_{j\mu }=0$)
and the "hadronic" sector with $A_{j\mu }\neq 0$ , which in some
approximation may look like bound states of the $N$ "quarks". It is worth to
mention that a "leptonic" soliton (\ref{e13}) with a Kerr-Newman asymptotic
form is fermion, because its gyromagnetic ratio\cite{CART1968},\cite%
{NEWM1973} is $g=2$ . Therefore we should not worry about the fermionic
character of the ordinary leptons. But we must be careful. The Kerr-Newman
solution cannot be considered as a soliton because it is singular at $r=0$.
The solitonic solutions must be everywhere regular!

3) Possible Existence of\ "Families"\ of\ "Leptons"\ and\ "Quarks": The
integrability conditions of the complex structure can be formulated in the
spinor formalism. They imply that both spinors $o^{A}$ and $\iota ^{A}$ of
the dyad satisfy the same PDE

\begin{equation}
\xi ^{A}\xi ^{B}\nabla _{AA^{\prime }}\ \xi _{B}=0  \label{e16}
\end{equation}%
where $\nabla _{AA^{\prime }}$ is the covariant derivative connected to the
vierbein $e_{a}^{\ \mu }$. One can show\cite{P-R1984} that both $o^{A}$ and $%
\iota ^{A}$ must satisfy the algebraic integrability condition

\begin{equation}
\Psi _{ABCD}\xi ^{A}\xi ^{B}\xi ^{C}\xi ^{D}=0  \label{e17}
\end{equation}%
Namely, they are principal directions of the Weyl spinor $\Psi _{ABCD}$.
Therefore a curved spacetime may admit a limited number of complex
structures, which are directly related to its principal null directions. In
the present case of integrable complex structures we have $\Psi _{0}=0=\Psi
_{4}$ and the corresponding structures are classified to the following four
cases\ \ 
\begin{equation}
\begin{array}{l}
Case\ I:\Psi _{1}\neq 0\ ,\ \Psi _{2}\neq 0\ ,\ \Psi _{3}\neq 0 \\ 
\\ 
Case\ II:\Psi _{1}\neq 0\ ,\ \Psi _{2}\neq 0\ ,\ \Psi _{3}=0 \\ 
\\ 
Case\ III:\Psi _{1}\neq 0\ ,\ \Psi _{2}=0\ ,\ \Psi _{3}=0 \\ 
\\ 
Case\ D:\Psi _{1}=0\ ,\ \Psi _{2}\neq 0\ ,\ \Psi _{3}=0 \\ 
\end{array}
\label{e18}
\end{equation}%
There is some evidence that only case D may admit static solitons\cite%
{RAG2008b}. Notice that there must be a certain correspondence between the
"leptonic" and "hadronic" families, because for each pure geometric soliton
("lepton") there must be N colored confined gauge field excitations
("quarks").

4) Complex structure admitting spacetime metrics are restrictions of Kaehler
metrics in $%
%TCIMACRO{\U{2102} }%
%BeginExpansion
\mathbb{C}
%EndExpansion
^{4}$.

The spacetimes which admit a complex structure (the "leptonic" sector of the
model) are four dimensional CR manifolds with codimension four\cite{BAOU}.
The starting point is the observation that, in every coordinate neighborhood
of the spacetime, the reality relations of the null tetrad combined with (%
\ref{e5}) imply the following conditions 

\begin{equation}
\begin{array}{l}
dz^{0}\wedge dz^{1}\wedge d\overline{z^{0}}\wedge d\overline{z^{1}}=0 \\ 
\\ 
dz^{\widetilde{0}}\wedge dz^{\widetilde{0}}\wedge d\overline{z^{0}}\wedge d%
\overline{z^{1}}=0 \\ 
\\ 
dz^{\widetilde{0}}\wedge dz^{\widetilde{0}}\wedge d\overline{z^{\widetilde{0}%
}}\wedge d\overline{z^{\widetilde{0}}}=0%
\end{array}
\label{e19}
\end{equation}%
for the structure coordinates $z^{b}\equiv (z^{\alpha },\;z^{\widetilde{%
\alpha }})$,\ $\alpha =0,\ 1$. Hence we may conclude that there are two real
functions $\Psi _{11}$ , $\Psi _{22}$ and a complex one $\Psi _{12}$,
defined in neighborhoods of $%
%TCIMACRO{\U{2102} }%
%BeginExpansion
\mathbb{C}
%EndExpansion
^{4}$, such that

\begin{equation}
\Psi _{11}(\overline{z^{\alpha }},z^{\alpha })=0\quad ,\quad \Psi
_{12}\left( \overline{z^{\alpha }},z^{\widetilde{\alpha }}\right) =0\quad
,\quad \Psi _{22}\left( \overline{z^{\widetilde{\alpha }}},z^{\widetilde{%
\alpha }}\right) =0  \label{e20}
\end{equation}%
Notice the special dependence of the defining functions on the structure
coordinates. These functions define totally real submanifolds of $%
%TCIMACRO{\U{2102} }%
%BeginExpansion
\mathbb{C}
%EndExpansion
^{4}$ which permit the model to use the powerful formalism of the CR
manifolds and the analytic domains. Therefore I will very briefly review
these mathematics\cite{BAOU}.

The four real generic conditions of the vanishing of a $2\times 2$ hermitian
matrix valued function%
\begin{equation}
\rho (z^{a},\overline{z^{a}})=%
\begin{pmatrix}
\rho _{11} & \rho _{12} \\ 
\overline{\rho _{12}} & \rho _{22}%
\end{pmatrix}%
=0  \label{e21}
\end{equation}%
of $%
%TCIMACRO{\U{2102} }%
%BeginExpansion
\mathbb{C}
%EndExpansion
^{4}$ locally determine a four dimensional CR manifold with codimension four%
\cite{BAOU}. In the interesting generic case, it is a maximally totally real
submanifold $M$ of $%
%TCIMACRO{\U{2102} }%
%BeginExpansion
\mathbb{C}
%EndExpansion
^{4}$. It is evident that the defining functions are not unique. Any
appropriate combination of them, which does not introduce additional points,
is equally good for the definition of the submanifold $M$ of $%
%TCIMACRO{\U{2102} }%
%BeginExpansion
\mathbb{C}
%EndExpansion
^{4}$. For every such condition and in the corresponding neighborhood of $%
%TCIMACRO{\U{2102} }%
%BeginExpansion
\mathbb{C}
%EndExpansion
^{4}$ we may define a kaehlerian metric, which turns out to be lorentzian on
the manifold $M$.

I consider the following Kaehler metric \ 
\begin{equation}
\begin{array}{l}
ds^{2}=2\tsum\limits_{a,b}\frac{\partial ^{2}(\det \rho )}{\partial
z^{a}\partial \overline{z^{b}}}dz^{a}d\overline{z^{b}} \\ 
\end{array}
\label{e23}
\end{equation}%
A straightforward calculation gives ($f_{a\overline{b}}=\frac{\partial
^{2}(\det \rho )}{\partial z^{a}\partial \overline{z^{b}}}$) \ 
\begin{equation}
\begin{array}{l}
f_{a\overline{b}}=\rho _{22}\frac{\partial ^{2}\rho _{11}}{\partial
z^{a}\partial \overline{z^{b}}}+\frac{\partial \rho _{11}}{\partial z^{a}}%
\frac{\partial \rho _{22}}{\partial \overline{z^{b}}}+\frac{\partial \rho
_{22}}{\partial z^{a}}\frac{\partial \rho _{11}}{\partial \overline{z^{b}}}%
+\rho _{11}\frac{\partial ^{2}\rho _{22}}{\partial z^{a}\partial \overline{%
z^{b}}}- \\ 
\qquad -\overline{\rho _{12}}\frac{\partial ^{2}\rho _{12}}{\partial
z^{a}\partial \overline{z^{b}}}-\frac{\partial \overline{\rho _{12}}}{%
\partial z^{a}}\frac{\partial \rho _{12}}{\partial \overline{z^{b}}}-\frac{%
\partial \rho _{12}}{\partial z^{a}}\frac{\partial \overline{\rho _{12}}}{%
\partial \overline{z^{b}}}-\rho _{12}\frac{\partial ^{2}\overline{\rho _{12}}%
}{\partial z^{a}\partial \overline{z^{b}}} \\ 
\end{array}
\label{e24}
\end{equation}%
On the surface ($\rho =0$) the metric takes the lorentzian form \ 
\begin{equation}
\begin{array}{l}
ds^{2}|_{M}=2(\frac{\partial \rho _{11}}{\partial z^{a}}\frac{\partial \rho
_{22}}{\partial \overline{z^{b}}}+\frac{\partial \rho _{22}}{\partial z^{a}}%
\frac{\partial \rho _{11}}{\partial \overline{z^{b}}}-\frac{\partial 
\overline{\rho _{12}}}{\partial z^{a}}\frac{\partial \rho _{12}}{\partial 
\overline{z^{b}}}-\frac{\partial \overline{\rho _{12}}}{\partial z^{a}}\frac{%
\partial \rho _{12}}{\partial \overline{z^{b}}})dz^{a}d\overline{z^{b}}= \\ 
\\ 
\qquad =2(\ell \otimes n-m\otimes \overline{m}) \\ 
\end{array}
\label{e25}
\end{equation}%
where $\ell ,n$ are real 1-forms and $m$ is complex on $M$ according to the
following definitions \ 
\begin{equation}
\begin{array}{l}
\ell =i\sqrt{2}\frac{\partial \rho _{11}}{\partial z^{a}}dz^{a}\quad ,\quad
n=i\sqrt{2}\frac{\partial \rho _{22}}{\partial z^{a}}dz^{a} \\ 
\\ 
m=i\sqrt{2}\frac{\partial \overline{\rho _{12}}}{\partial z^{a}}dz^{a}\quad
,\quad \overline{m}=i\sqrt{2}\frac{\partial \rho _{12}}{\partial z^{a}}dz^{a}
\\ 
\end{array}
\label{e26}
\end{equation}%
because $d\rho |_{M}=0$ on the real submanifold. Notice that a different
defining function changes the Kaehler metric in the corresponding
neighborhood of $%
%TCIMACRO{\U{2102} }%
%BeginExpansion
\mathbb{C}
%EndExpansion
^{4}$ .

Using the implicit function theorem, the defining relations $\rho =0$ may
take\cite{BAOU} the general form $y^{a}=h^{a}(x^{b})$ where $x^{a}=\func{Re}%
(z^{a})$ and $y^{a}=\func{Im}(z^{a})$. If the defining functions are real
analytic at any point of $M$, the real submanifold is CR (holomorphically)
equivalent to the simple case $y^{\prime a}=0$, which may take the following
hermitian matrix form 
\begin{equation}
\begin{array}{l}
\rho =\frac{i}{\sqrt{2}}%
\begin{pmatrix}
(\overline{r^{0}}-r^{0})-(\overline{r^{3}}-r^{3}) & -(\overline{r^{1}}%
-r^{1})+i(\overline{r^{2}}-r^{2}) \\ 
-(\overline{r^{1}}-r^{1})-i(\overline{r^{2}}-r^{2}) & (\overline{r^{0}}%
-r^{0})+(\overline{r^{3}}-r^{3})%
\end{pmatrix}
\\ 
\end{array}
\label{e27}
\end{equation}%
with $r^{a}=x^{\prime a}+iy^{\prime a}$. Then the implied metric takes the
form \ 
\begin{equation}
\begin{array}{l}
ds^{2}=\frac{1}{2}\tsum\limits_{a,b}\frac{\partial ^{2}(-(\overline{r^{c}}%
-r^{c})^{2})}{\partial r^{a}\partial \overline{r^{b}}}dr^{a}d\overline{r^{b}}%
=\eta _{ab}dr^{a}d\overline{r^{b}} \\ 
\end{array}
\label{e28}
\end{equation}%
which apparently becomes the Minkowski metric on the surface $y^{\prime a}=%
\func{Im}(r^{a})=0$. If there is at least one point of $M$, where its
defining function $h^{a}(x^{b})$ is not real analytic, the corresponding
Kaehler metric cannot be reduced down to the Minkowski metric. Therefore we
conclude that at least at the metric level only non real analytic CR
submanifolds contain gravity.

In the present case of totally real submanifolds of $%
%TCIMACRO{\U{2102} }%
%BeginExpansion
\mathbb{C}
%EndExpansion
^{4}$\ which admit a defining condition with the precise (\ref{e19}) $z^{a}$
dependence, the null tetrad (\ref{e26}) is integrable and the defined metric
admits a complex structure. Hence we conclude that this kind of spacetime
metrics can always be extended into a Kaehler metric in a neighborhood of $%
%TCIMACRO{\U{2102} }%
%BeginExpansion
\mathbb{C}
%EndExpansion
^{4}$. In this case the lorentzian complex structure preserving
transformations are $z^{\alpha }=f^{\alpha }(z^{\beta })$ and $z^{\widetilde{%
\alpha }}=f^{\widetilde{\alpha }}(z^{\widetilde{\beta }})$ and not the
general holomorphic transformations of $%
%TCIMACRO{\U{2102} }%
%BeginExpansion
\mathbb{C}
%EndExpansion
^{4}$\ considered above.

It is already known that any hypersurface type (codimension-1) CR manifold,
extended with a line bundle, admits a Kaehler metric\cite{FE}. It would be
interesting to investigate whether the present Kaehler metric has any
relation with the asymptotic expansion of the Bergman kernel at the boundary
of the corresponding domain 
\begin{equation}
\rho (z^{a},\overline{z^{a}})=%
\begin{pmatrix}
\rho _{11} & \rho _{12} \\ 
\overline{\rho _{12}} & \rho _{22}%
\end{pmatrix}%
>0  \label{28}
\end{equation}%
analogous to the relation of the ordinary Fefferman metric.

We are now formally ready to prove the emergence of the Poincar\'{e}
symmetry, which opens up possible physical relevance of the present model.
The Poincar\'{e} group must be an exact symmetry in the "particle" spectrum
of the model for it to have any physical relevance. Recall that in General
Relativity the symmetry of asymptotically flat spacetimes is the BMS group,
which does not appear in nature.

5) The "Leptonic" Sector Belongs to Representations of the Poincar\'{e}
Group: The Poincar\'{e} group is an exact symmetry of the present model
because it preserves the lorentzian complex structure and the vacuum of the
model\cite{RAG2008b}.

In the mathematical study of CR manifolds, an essential step is to osculate
the real surface with the boundary of a classical domain. In the simple case
of hypersurface type CR manifolds the hyperquadric is used\cite{JACO} to
osculate them. It is apparent\cite{RAG2008b} that in the present case the
most convenient real surface is the $U(2)$ characteristic boundary of the $%
SU(2,2)$ invariant classical domain. The mathematical study of this kind of
problems is performed after their projective formulation. For this purpose I
consider the rank-2 $4\times 2$ matrices $X^{mi}$\ with every column being a
point of an algebraic surface $K_{i}(X^{mi})$ of the $CP^{3}$\ projective
space, which Penrose calls twistor. Then I consider that the $2\times 2$
matrix $\Psi $ has the form%
\begin{equation}
\Psi =X^{\dagger }EX-%
\begin{pmatrix}
G_{11} & G_{12} \\ 
\overline{G_{12}} & G_{22}%
\end{pmatrix}
\label{e29}
\end{equation}%
where $E$\ is an $SU(2,2)$ invariant $4\times 4$ matrix and $G_{ij}=G_{ij}(%
\overline{X^{mi}},X^{mj})$ are homogeneous functions. In the simple case $%
G_{ij}=0$ it is the boundary of the first kind Siegel domain\cite{PYAT} for
\ 
\begin{equation}
E=%
\begin{pmatrix}
0 & I \\ 
I & 0%
\end{pmatrix}
\label{e30}
\end{equation}%
which is holomorphic to the $SU(2,2)$ invariant bounded classical domain
given by \ 
\begin{equation}
E=%
\begin{pmatrix}
I & 0 \\ 
0 & -I%
\end{pmatrix}
\label{e31}
\end{equation}

Using the following spinorial form of the rank-2 matrix $X^{mj}$ in its
unbounded realization 
\begin{equation}
\begin{array}{l}
X^{mj}=%
\begin{pmatrix}
\lambda ^{Aj} \\ 
-ir_{A^{\prime }B}\lambda ^{Bj}%
\end{pmatrix}
\\ 
\end{array}
\label{e32}
\end{equation}%
and the null tetrad 
\begin{equation}
\begin{array}{l}
L^{a}=\frac{1}{\sqrt{2}}\overline{\lambda }^{A^{\prime }1}\lambda
^{B1}\sigma _{A^{\prime }B}^{a}\quad ,\quad N^{a}=\frac{1}{\sqrt{2}}%
\overline{\lambda }^{A^{\prime }2}\lambda ^{B2}\sigma _{A^{\prime
}B}^{a}\quad ,\quad M^{a}=\frac{1}{\sqrt{2}}\overline{\lambda }^{A^{\prime
}2}\lambda ^{B1}\sigma _{A^{\prime }B}^{a} \\ 
\\ 
\epsilon _{AB}\lambda ^{A1}\lambda ^{B2}=1 \\ 
\end{array}
\label{e33}
\end{equation}%
we find\cite{RAG2011} \ 
\begin{equation}
\begin{array}{l}
y^{a}=\frac{1}{2\sqrt{2}}[G_{22}N^{a}+G_{11}L^{a}-G_{12}M^{a}-\overline{%
G_{12}}\overline{M}^{a}] \\ 
\end{array}
\label{e34}
\end{equation}%
where $y^{a}$\ is the imaginary part of $r^{a}=x^{a}+iy^{a}$\ defined by the
relation $r_{A^{\prime }B}=r^{a}\sigma _{aA^{\prime }B}$\ . The matrices $%
\sigma _{A^{\prime }B}^{a}$ being the identity and the three Pauli matrices.
If we substitute the normalized $\lambda ^{Ai}$\ as functions of $r^{a}$,
using the Kerr conditions $K_{i}(X^{mi})$, these relations turn out to be
four real functions of $x^{a}$ and $y^{a}$. The implicit function theorem
assures the existence of the solution $y^{a}=$ $h^{a}(x)$ which we have also
found above. If the manifold is real analytic, an ordinary holomorphic
transformation (which does not preserve the Flaherty lorentzian complex
structure) will make the induced metric conformally flat. Therefore I will
consider real submanifolds with a non-real analytic point transferred at
infinity.

Notice that a surface (\ref{e34}) does not generally belong into the Seigel
domain, because $y^{0}$\ and \ 
\begin{equation}
\begin{array}{l}
y^{a}y^{b}\eta _{ab}=\frac{1}{8}[G_{22}G_{11}-G_{12}\overline{G_{12}}] \\ 
\end{array}
\label{e35}
\end{equation}%
are not always positive. But the regular surfaces (with an upper bound) can
always be brought inside the Siegel domain (and its holomorphic bounded
classical domain) with an holomorphic complex time translation 
\begin{equation}
\begin{array}{l}
X^{\prime }=%
\begin{pmatrix}
I & 0 \\ 
dI & I%
\end{pmatrix}%
X \\ 
\end{array}
\label{e36}
\end{equation}%
of (\ref{e32}), where $d$ is a real constant large enough to transfer the
manifold inside the Siegel classical domain, because it implies $y^{^{\prime
}0}=y^{0}+id$. This transformation preserves the point at infinity.
Therefore we can always assume that the CR manifold is always inside the $%
SU(2,2)$ invariant classical domain, up to its point at the boundary. The $%
SU(2,2)$ is a symmetry of the solution submanifolds which are inside the
classical domain, because it preserves the lorentzian complex structure and
the classical domain.

The existence of an irregular point (infinity) at the characteristic
boundary, breaks the $SU(2,2)$ symmetry of the pure geometric solutions down
to its Poincar\'{e}$\times $dilation subgroup, which preserves\cite{PYAT}
the infinity point being on the Shilov boundary. The vacuum of the model
must be an open real submanifold with a precise lorentzian complex
structure, preserved by the Poincar\'{e} subgroup. It is related with the
definition of a positive conserved quantity which could play the role of the
"energy" of the model. The great success of the Einstein equations suggests
that this quantity should be related with the Einstein tensor of the
spacetime. If we could prove it, the Einstein equations will turn out to be
definitions of the excitation modes of the lorentzian complex structure.
While the functional minima of the "energy" will be identified with the
"leptons" of the model\cite{RAG2008b}.

The breaking of the dilation subgroup seems to be related with this "energy"
quantity and possibly with the non-real-analytic property of infinity.
Notice that the asymptotic form of the potential of a massive field $\frac{%
e^{-mr}}{r}$ is a typical function which is not real analytic at $r=\infty $%
. 

Concluding the presentation of the model, I want to point out that the exact
Poincar\'{e} symmetry assures the phenomenological description of the
"particle" interactions (like the Standard Model) using ordinary Quantum
Fields with the ordinary quadratic terms, because simply these fields are
exactly the Poincar\'{e} representations of the "particles". I have not yet
found the explicit forms of the lorentzian complex structure modes, the
"leptonic" solitons and the "hadronic" bound states of the model, but it is
apparent that the general picture indicates that it may have some physical
relevance.

\bigskip

\end{document}